\documentclass{jfm}
\usepackage{graphicx}
\usepackage{newtxtext}
\usepackage{newtxmath}
\usepackage{natbib}
\usepackage{hyperref}
\usepackage{ulem}
\hypersetup{
colorlinks = true,
urlcolor = blue,
citecolor = black,
}

\newcommand{\RomanNumeralCaps}[1]
\linenumbers

\usepackage{xcolor}

\def\tb{\textcolor{black}}

\begin{document}

\shorttitle{Pressure fluctuations of liquids under short-time acceleration} 
\shortauthor{C. Kurihara et al.} 

\title{Pressure fluctuations of liquids under short-time acceleration}

\author
{
Chihiro Kurihara\aff{1},
Akihito Kiyama\aff{2,3},
\and 
Yoshiyuki Tagawa\aff{1,2}
\corresp{\email{tagawayo@cc.tuat.ac.jp}}
}

\affiliation
{\aff{1}
Department of Mechanical Systems Engineering, Tokyo University of Agriculture and Technology, Nakacho 2-24-16 Koganei, Tokyo 184-8588, Japan.
\aff{2}
Institute of Global Innovation Research, Tokyo University of Agriculture and Technology, Nakacho 2-24-16 Koganei, Tokyo 184-8588, Japan.
\aff{3}
Department of Mechanical Science, Saitama University, Saitama, Japan.
}

\maketitle
\begin{abstract}
{This study experimentally investigates the pressure fluctuations of liquids in a column under short-time acceleration and demonstrates that the Strouhal number $St$ [$=L/(c\Delta t)$, where $L$, $c$, and $\Delta t$ are the liquid column length, speed of sound, and acceleration duration, respectively] provides a measure of the pressure fluctuations both for limiting cases (i.e. $St\ll1$ or $St = \infty$) and for intermediate $St$ values. 
Incompressible fluid theory and water hammer theory respectively imply that the magnitude of the averaged pressure fluctuation $\overline{P}$ becomes negligible for $St\ll1$ (i.e., in the condition where the duration of acceleration $\Delta t$ is large enough compared to the acoustic timescale) and tends to $\rho cu_0$ (where $u_0$ is the change in the liquid velocity) for $St\geq O(1)$ (i.e., in the condition where $\Delta t$ is small enough). 
For intermediate $St$ values, there is no consensus on the value of $\overline{P}$. 
In our experiments, $L$, $c$, and $\Delta t$ are varied so that $0.02 \leq St \leq 2.2$.
The results suggest that the incompressible fluid theory holds only up to $St\sim0.2$ and that $St$ governs the pressure fluctuations under different experimental conditions for higher $St$ values.
The data relating to a hydrogel also tend to collapse to a unified trend.
The inception of cavitation in the liquid starts at $St\sim 0.2$ for various $\Delta t$, indicating that the liquid pressure becomes negative.
To understand this mechanism, we employ a one-dimensional wave propagation model with a pressure wavefront of finite thickness that scales with $\Delta t$.
The model provides a reasonable description of the experimental results as a function of $St$.
The slight discrepancy between the model and experimental results reveals additional contributing factors such as the container motion and the profile of the pressure wavefront.
}

\end{abstract}

\section{Introduction}
Abrupt liquid motion causes pressure fluctuations inside the liquid column, resulting in significant fluid motion.
For example, a large deformation of the free surface (i.e. jets) can occur in Pokrovski's experiment, where a liquid-filled container falls \tb{under gravity} and hits the floor \citep{Antkowiak2007}.
The pressure impulse approach under the incompressible fluid assumption explains the jet formation.
Nevertheless, \cite{Kiyama2016} found that sufficiently large acceleration causes cavitation accompanied by jet formation with a vibrating interface. 
This implies that the acoustic pressure waves propagate in the liquid column \citep{Bao2023}, meaning that the jet liquid behaves as a weakly compressible fluid. 

To describe the liquid compressibility, scaling analysis \citep[\S 6.3, p. 168-]{Batchelor1967} provides three important terms related as
\begin{equation}
\left|\frac{1}{\rho c^2}\frac{\partial p}{\partial t}-\frac{1}{2c^2}\frac{Dq^2}{Dt}+\frac{aU}{c^2}\right|\ll\frac{U}{L},
\end{equation}
where $\rho$ is the liquid density, $c$ is the speed of sound, $p$ is the pressure, $q$ is a quantity having the same dimensions as velocity, $a$ is the typical acceleration, $U$ is the typical velocity of liquid (e.g., the impact velocity in an example of Pokrovski's experiment), and $L$ is the typical length scale of the liquid system. 
Note that the flow is assumed to be isentropic.
When each term on the left-hand side has a much smaller magnitude than the spatial derivatives of the components of liquid velocity $U/L$, the liquid behaves as if it were incompressible.
\tb{This consideration gives} three dimensionless \tb{incompressibility conditions} relating to the Mach number $Ma$, Strouhal number $St$, and Froude number $Fr$, defined as 
\begin{equation}
\frac{L}{c\Delta t}=St\ll1,
\end{equation}
\begin{equation}
\frac{U}{c}=Ma\ll1,
\end{equation}
\begin{equation}
\frac{aL}{c^2}=Fr^{-1}\ll1,
\end{equation}
where $\Delta t$ is the duration required for the development of the pressure field.
The Strouhal number $St$ is the ratio of the duration of acceleration $\Delta t$ to the acoustic timescale $L/c$; a detailed explanation will be provided in \S 3.1.
Note that the Froude number $Fr$ is generally expressed as the ratio of the inertial force to gravitational force.
The physical meaning of $Fr$ in the above equation is the competition between the pressure change due to the body force (acceleration including gravity) $\sim\rho aL$ and the ambient pressure $\sim\rho c^2$.
This dimensionless number $aL/c^2$ is equivalent to $Fr^{-1}$, and is also the product of the other two dimensionless numbers (i.e. $Fr^{-1}\sim StMa$) when the typical acceleration can be scaled as $a\sim U/\Delta t$. 

In a \tb{previous} example of Pokrovski's experiment with fluid compressibility effects \citep{Kiyama2016}, the typical values were found to be $U\sim O(10^0)$ m/s, $c\sim O(10^3)$ m/s, $L\sim O(10^{-1})$ m, $\Delta t\sim O(10^{-4})$ s, and $a\sim O(10^4)$ m/s$^2$. 
Under such conditions, $Ma\sim O(10^{-3})\ll1$, $St\sim O(10^0)$, and $Fr^{-1}\sim O(10^{-3})\ll1$.
\tb{This suggests} that the compressibility effect in Pokrovski's experiment might be scaled with the Strouhal number $St$, rather than with $Ma$ and $Fr$.
\tb{However, \citet{Kiyama2016} did not consider the influence of the acceleration duration (i.e. $St$).}

\begin{figure}
\begin{center}
\includegraphics[scale=0.3]{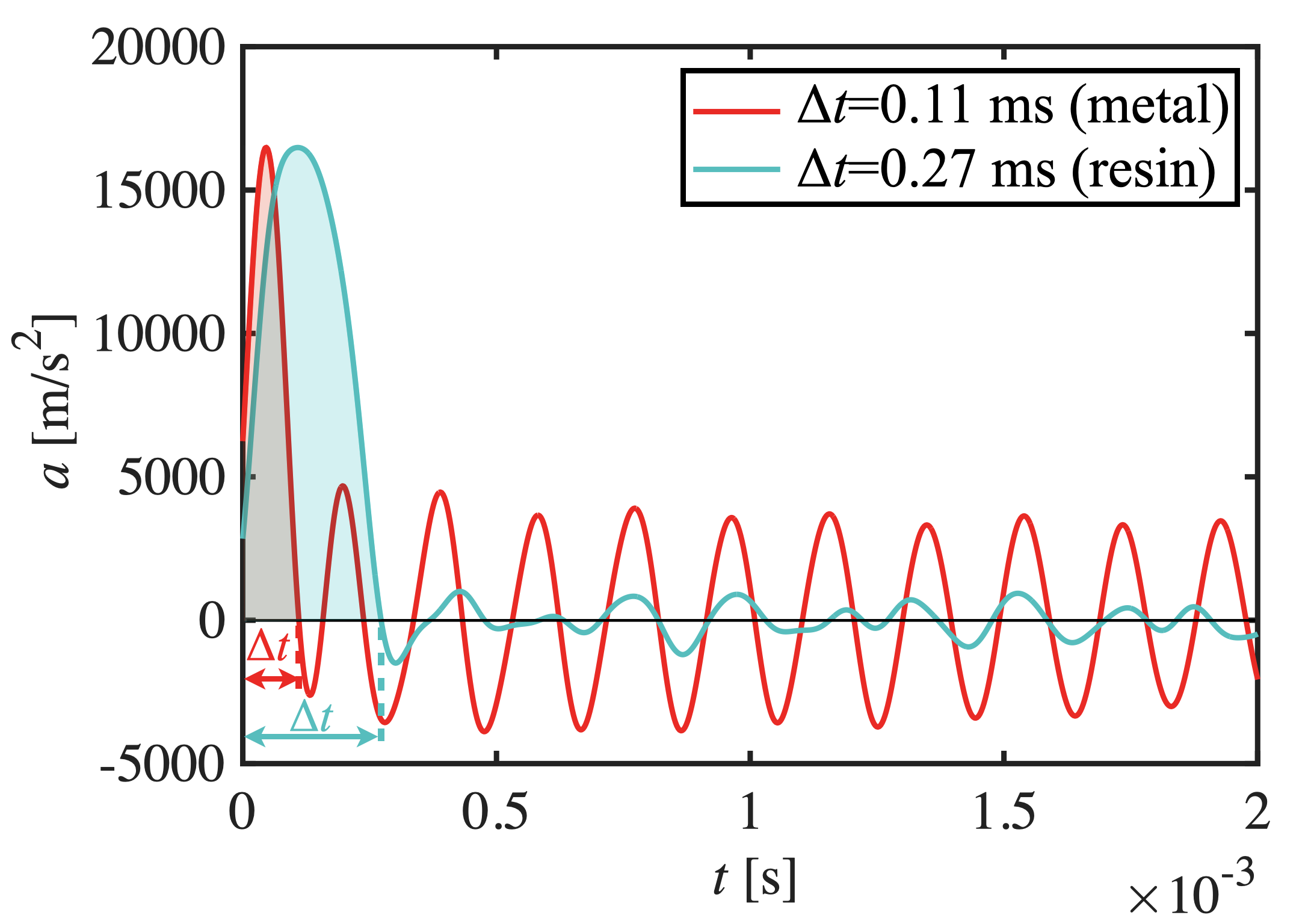}
\caption{Measured acceleration of a \tb{liquid-filled} glass container after \tb{collision} with the floor. Red and blue curves show the acceleration \tb{with} different \tb{floor} materials \tb{and drop heights}. The acceleration during the impact is \tb{marked} by \tb{the} shaded area. Although both cases have similar peaks and mean accelerations at impact, the magnitude of the subsequent acceleration fluctuations is quite different.}
\label{fig:Compare}
\end{center}
\end{figure}

\begin{figure}
\begin{center}
\includegraphics[scale=0.4]{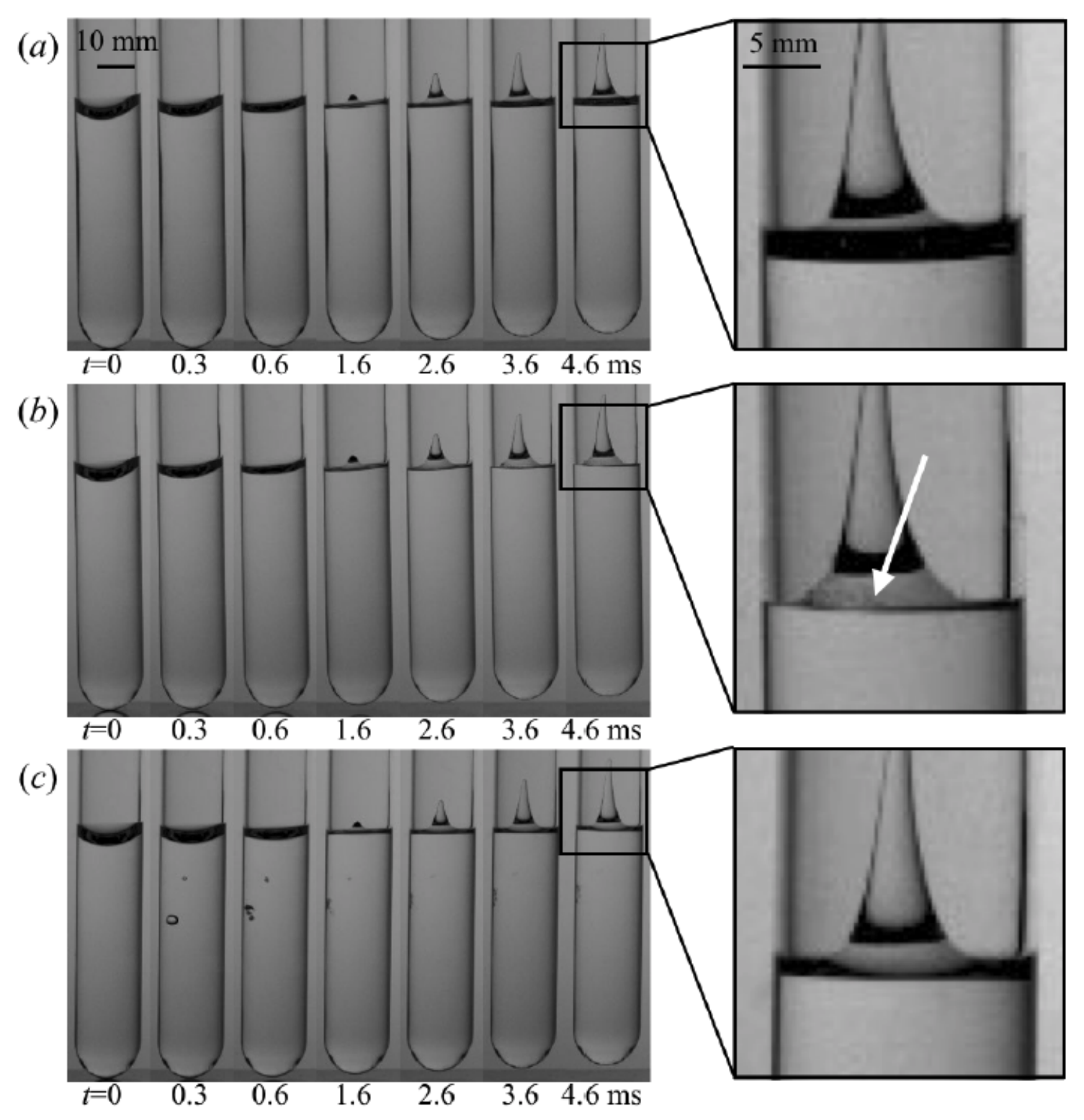}
\caption{Photographs taken at 100,000 fps using a high-speed camera (Photron SA-X) with a back-light method. The typical velocity is mostly constant ($\approx$ 2.0 m/s) and the height of the jet does not \tb{change} significantly in each case. The rightmost panels show a magnified view of the free surface at $t=$4.6 ms. ({\it a}) Test tube impacting a resin floor, where the free surface remains smooth. ({\it b}) Test tube impacting a metal floor, where the free surface vibrates periodically and exhibits a rough texture \tb{(marked by a white arrow)}. ({\it c}) Test tube impacting a metal floor, where cavitation occurs inside the liquid. The free surface exhibits a similar response to that in ({\it a}). See supplementary movies.}
\label{fig:images}
\end{center}
\end{figure}

We \tb{tested the above hypothesis in} a preliminary experiment. 
Figure \ref{fig:Compare} shows the time series of the acceleration $a$ measured at the top of a container filled with \tb{silicone oil (10~cSt)} as it impacts the floor.
The red line indicates the data taken from the tube impacting a metal floor, where the duration of acceleration $\Delta t\sim$0.11~ms (indicated by the red shading).
The blue line presents the data for a resin floor ($\Delta t\sim$0.27~ms, blue shading).
\tb{The dimensionless parameters are $St\sim 3.1\times10^{-1}$, $Ma\sim 8.2\times10^{-4}\ll1$, and $Fr^{-1}\sim 3.6\times10^{-5}\ll1$ (red) and $St\sim 1.2\times10^{-1}$, $Ma\sim 2.1\times10^{-3}\ll1$, and $Fr^{-1}\sim 3.6\times10^{-5}\ll1$ (blue).}
Although both cases exhibit similar peaks and mean accelerations at impact, the magnitude of later acceleration fluctuations is quite different.
\tb{The case with the larger Strouhal number $St$ (i.e. a smaller $\Delta t$) exhibits} more significant acceleration fluctuations for $t>\Delta t$.
This \tb{confirms} that the \tb{short-term} acceleration \tb{induces} pressure waves in the liquid column, even when $Ma\ll 1$ and $Fr^{-1}\ll 1$.
We conjectured that the influence of $\Delta t$ (and thus $St$) became visible there, as the other two dimensionless parameters ($Ma, Fr^{-1}$) remain much smaller than unity, implying that they are less important.
The effects of the acceleration duration $\Delta t$ are \tb{also visible in} the fluid motion inside test tubes dropped from the same height (i.e. the same velocity $U$, see figure \ref{fig:images}).
\tb{The resin floor induces a smooth jet (figure \ref{fig:images}{\it a}), while the metal floor triggers surface vibrations (figure \ref{fig:images}{\it b}) and cavitation (figure \ref{fig:images}{\it c})}; see supplementary movies.

To the best of our knowledge, the role of $\Delta t$ in the development of the pressure field in Pokrovski's experiment has not been systematically investigated. 
Most existing research took a pressure impulse approach assuming an incompressible flow and a constant $\Delta t$ which is greater than the acoustic timescale $L/c$, i.e. $St\ll1$ \citep{Antkowiak2007,Pan2017}.
In this approach, the pressure field of the liquid is fully developed immediately after the impact, and thus the liquid pressure does not change as a function of time.
Another approach takes the water hammer theory into account \citep{Kiyama2016}, and thus assumes an instant increment in pressure, i.e. $\Delta t=0$ \citep{Ghidaoui2005,Bergant2006}, \tb{where $St=\infty$}.
This assumption predicts periodic pressure fluctuations $\overline{P}$ with a magnitude of $\rho cU$, but does not capture the role of $\Delta t$ in the development of the pressure field.
Existing research has considered one or other of these approaches, but the intermediate region between $St\ll1$ and $St=\infty$ has rarely been studied.

This paper focuses on the role of $St$ in the pressure fluctuations, especially in the intermediate $St$ regime.
We first define the physical meaning of $St$ in \S 3.1 as a function of $\Delta t$, and examine the experimental data with \tb{various} liquid depths $L$, velocities $U$, acceleration durations $\Delta t$, and liquid types.
The experimental data collapse onto a single curve, suggesting that $St$ is suitable for describing the pressure fluctuations (\S 3.2).
\tb{The conditions for the onset of cavitation are derived in \S 3.2.2 and the pressure fluctuations in a hydrogel are examined in \S 3.2.3. In both cases, the present results are shown to be in line with existing pressure results for fluids.}
We also develop a simple model which takes the effect of the finite thickness of the pressure wavefront into account. This model is used to describe the liquid pressure fluctuations at $t>\Delta t$ based on the one-dimensional wave equation (\S 3.3).
We then compare the output from the proposed model with a wide range of experimental data.
The influence of the motion of the surrounding container and the profile of the pressure wavefront is also discussed.

\section{Experiments}
\begin{figure}
\begin{center}
\includegraphics[scale=0.15]{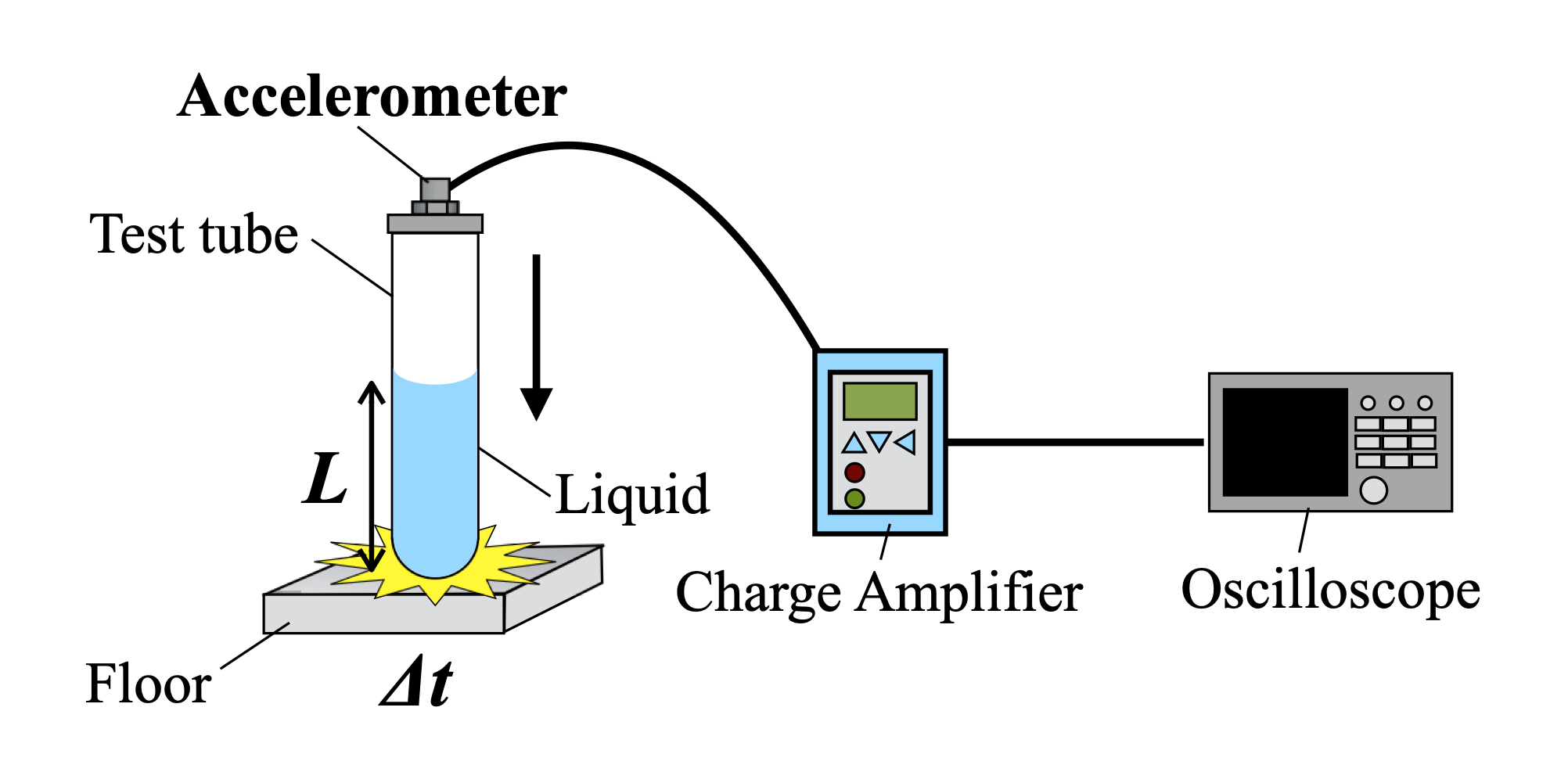}
\caption{Schematic illustration of the experimental setup. The accelerometer attached to the top of the container records the acceleration $a$ acting on the system. A change in the floor material allows us to change the duration of acceleration $\Delta t$. The depth of the liquid in the container $L$ is also varied in our experiments.}
\label{fig:Setup}
\end{center}
\end{figure}

Figure \ref{fig:Setup} shows \tb{the} experimental setup.
The container, which is partially filled with a liquid\tb{/hydrogel}, falls freely and eventually collides with the floor.
This accelerates the liquid in the vertical direction opposite to the direction of gravity $g$.
The acceleration of the container is measured with an accelerometer [2350, Showa Sokki Co., sensibility 0.3 pC/(m/s$^{^2}$)] attached to the top of \tb{the} container.
The accelerometer outputs a charge which is converted to a voltage by a charge amplifier (5015A, 5011B, Kistler Co.), and 
the voltage is recorded by an oscilloscope (Iwatsu Co., ViewGo II, DS 5554-A).

The control parameters are the depth of the liquid column $L$, the change in the liquid velocity induced by the impact $u_{0}$, the duration of acceleration $\Delta t$, and the type of liquid/hydrogel.
The depth of the liquid column $L$ ranges from 30--285~mm.
We use two borosilicate glass containers \tb{of different sizes and masses} (see table \ref{table:test tube}).
The characteristic velocity $U$ in this experiment corresponds to the impact velocity $u_0$\tb{, which} is varied from 0.5--3~m/s by \tb{adjusting} the drop height of the container.
The duration of acceleration $\Delta t$ ranges from 0.1--2.2~ms.
To control $\Delta t$, different floor materials are \tb{used}: steel, aluminium, epoxy resin, ABS resin, and rubber.
\tb{We} use different fluids (silicone oils with 1 and 10~cSt obtained from Shin-Etsu Chemical Co., pure water, and ethanol) and a weak gelatin gel (5 wt\%) that is expected to flow when it experiences rapid deformation \citep{Kiyama2019} and exhibits similar cavitation to water \citep{Rapet2019}. 
The physical properties of these media are summarized in table \ref{table:Liquids}.

\begin{figure}
\begin{center}
\includegraphics[scale=0.065]{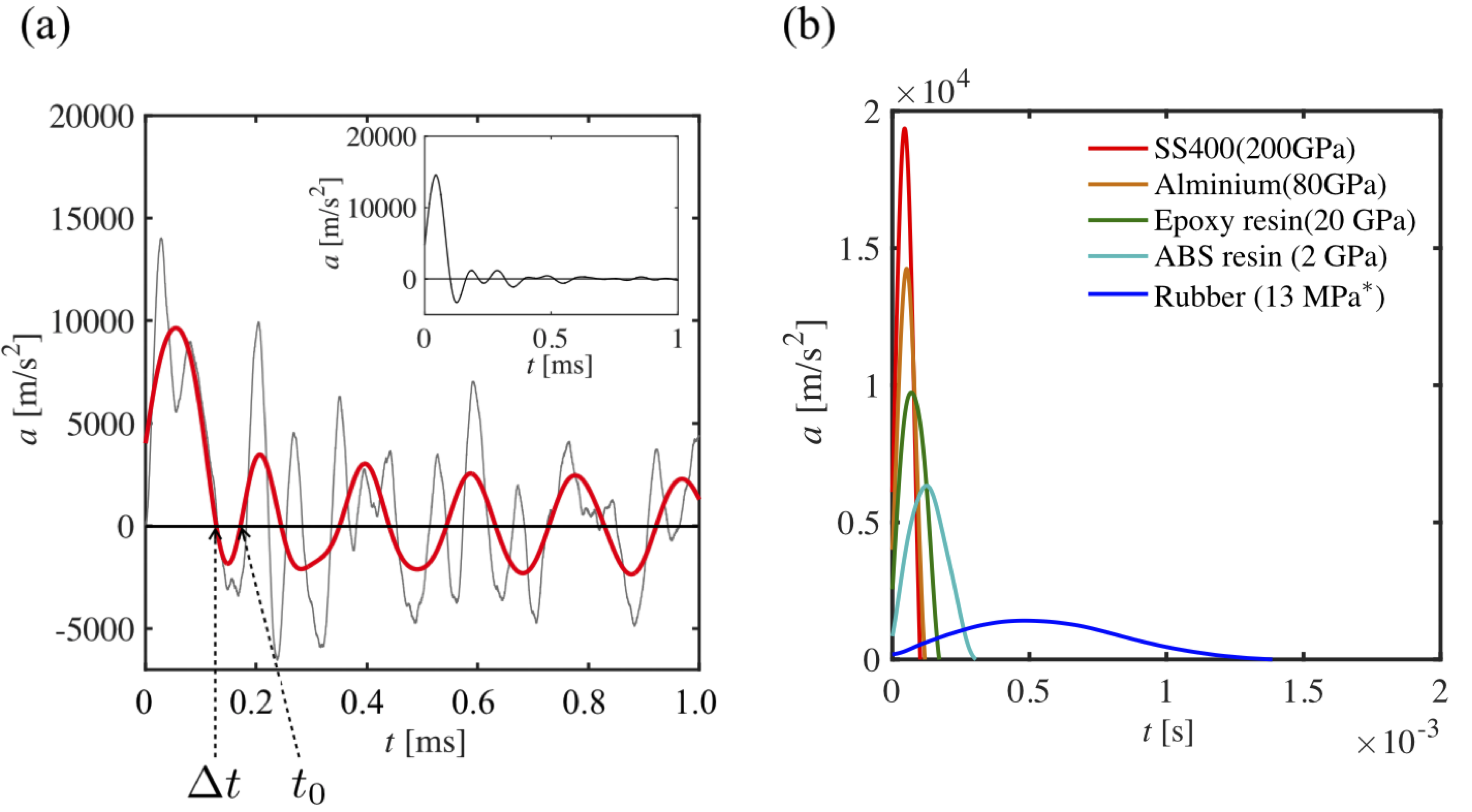}
\caption{({\it a}) Acceleration of the container partially filled with silicone oil (10~cSt). The grey and red lines represent raw data and data subjected to a low-pass filter. $\Delta t$ is the duration of acceleration. After $t_{0}$ (defined in \S 3.4), \tb{fluctuations in acceleration are visible}. The inset shows the acceleration of an empty glass container. ({\it b}) Acceleration of each \tb{floor type is} shown by different colours. Data are obtained from the impact of an empty test tube on each floor material. The Young's modulus is specified in the legend, \tb{and $^\ast$ denotes the tensile strength}.}
\label{fig:Acceleration}
\end{center}
\end{figure}

\tb{A} measured acceleration profile is shown in figure \ref{fig:Acceleration}({\tb{\it a}}).
A positive value of the acceleration indicates upward vertical acceleration.
The grey curve shows the raw acceleration data, \tb{whereas the red curve shows the low-pass-filtered data}. 
\tb{For filtering,} the cut-off frequencies for the 130 and 300~mm tube lengths are 11,000 and 5,000~Hz, respectively; these values are \tb{estimated} based on the tube length and the speed of sound in glass \tb{($c\sim 5.4\times10^3$~m/s)}.
The raw data suddenly change at $t=0$ when the container \tb{impacts} the floor, and the acceleration exceeds 1,000 $g$ immediately after the impact.
We define $\Delta t$ as the \tb{duration} from the collision until the filtered acceleration falls back to 0~m/s$^2$. 
Periodic fluctuations in acceleration are \tb{then} observed (see figure \ref{fig:Acceleration} for $t >t_{0}$).
\tb{At $t>t_0$, no fluctuations are visible for an empty container (see inset), indicating that there must be some liquid/hydrogel in the container to observe fluctuations.}

The acceleration during $\Delta t$ for each floor \tb{type} is shown in figure \ref{fig:Acceleration}({\it b}).
The container is dropped from the same height \tb{so that} the integral of the acceleration over $\Delta t$ is similar for all cases.
\tb{However}, a stiffer floor introduces a higher magnitude of acceleration $a$ \tb{within a} shorter $\Delta t$.

\begin{figure}
\begin{center}
\includegraphics[scale=0.14]{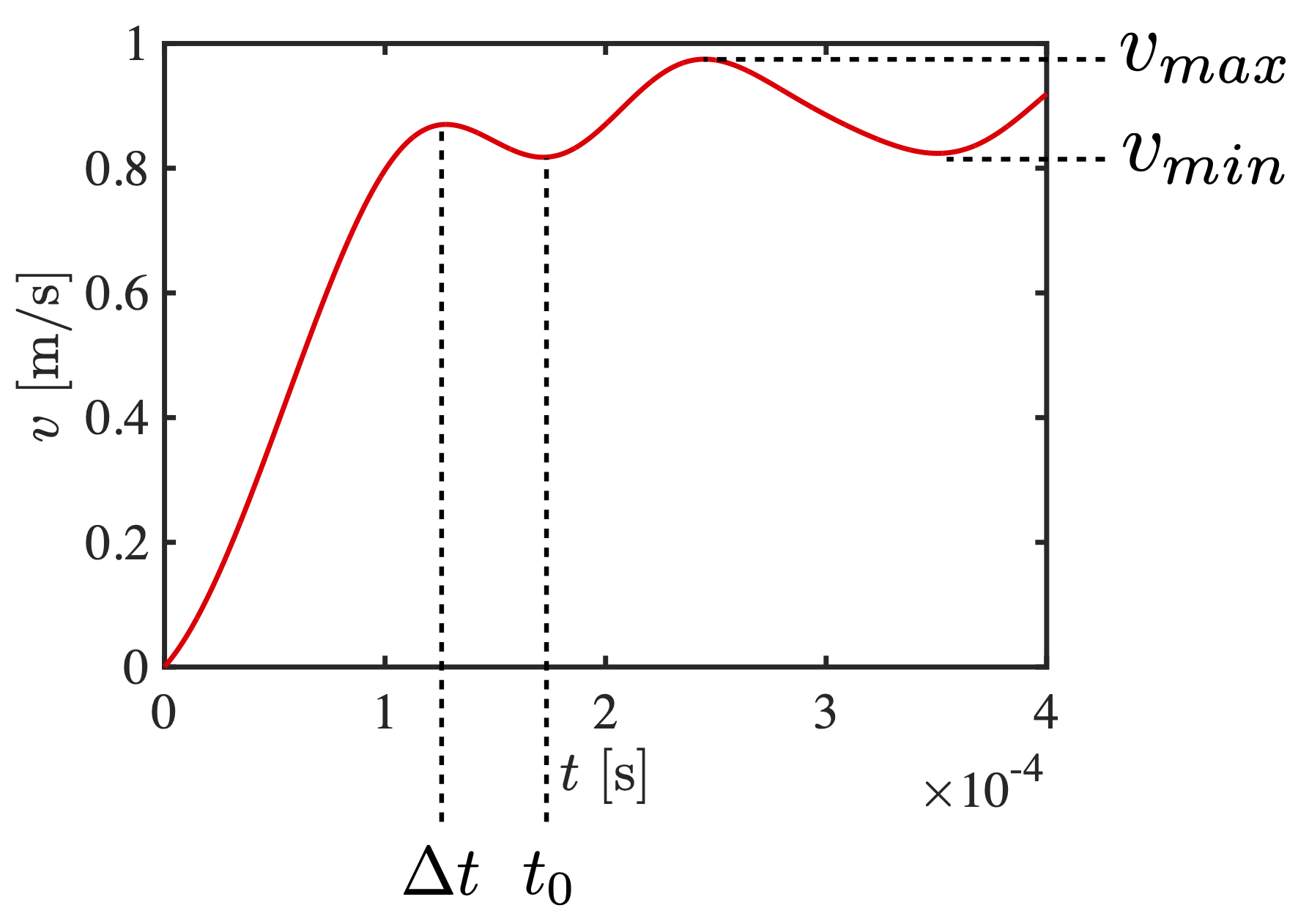}
\caption{Temporal velocity of the container (\tb{i.e.} test tube) from the beginning of impact, \tb{as calculated by integrating the acceleration data}. The peak velocities $v_{max}$ and $v_{min}$ are \tb{marked}.}
\label{fig:Velocity}
\end{center}
\end{figure}

The measured acceleration is translated to pressure according to the momentum conversation law: the amplitude of momentum change in the total liquid, $\Pi_{L}$, is equal to that of the container, $\Pi_{T}$.
The amplitude of the momentum change in the total liquid is expressed as 
\begin{equation}
\Pi_{L}=\rho A \frac{(\int_{0}^{L}udx)_{max}-(\int_{0}^{L}udx)_{min}}{2}=\rho A L \overline{U},
\label{Liquid_momentum} \end{equation}
where $\overline{U}$ represents the amplitude of the time difference and the spatial average of the velocity of the liquid, and $A$ indicates the cross-sectional area of the liquid column.
The amplitude of the change in momentum of the container is expressed as 
\begin{equation}
\Pi_{T}=mV,
\label{Test_tube_momentum} \end{equation}
where $m$ and $V$ are the mass of the container (see table 1) and the amplitude of the velocity fluctuations of the container, respectively.
\tb{Coupling} equations (\ref{Liquid_momentum}) and (\ref{Test_tube_momentum}) gives the following expression:
\begin{equation}
\frac{\overline{U}}{u_{0}}=\frac{m}{\rho A L}\frac{V}{u_0}.
\label{eq:comparison}
\end{equation}
\tb{We can experimentally determine} the right-hand side of equation (\ref{eq:comparison}).
We \tb{estimate} the velocity of the container $V$ by integrating the acceleration data (see figure \ref{fig:Acceleration}{\it a}).
We focus on the velocity after $t = t_{0}$ and detect the velocity fluctuations $v_{max}$ and $v_{min}$ (see figure \ref{fig:Velocity}).
\tb{Note that} $v_{max}$ and $v_{min}$ are \tb{taken from} the first period after $t=t_{0}$ \tb{to reduce the influence of} drift noise.
The magnitude of the fluctuations in $V$ is defined as the mean value of $v_{max}$ and $v_{min}$.
\tb{Assuming the relationship $\overline{P}=\rho c\overline{U}$, we obtain}
\begin{equation}
\frac{\overline{U}}{u_{0}}=\frac{\overline{P}}{\rho c u_{0}}.
\label{eq:iikae}
\end{equation}
Hereafter, the experimental data \tb{$\overline{P}/(\rho cu_0)$} are \tb{estimated as} the representation of $mV/(\rho ALu_0)$.

\begin{table}
\begin{center}
\begin{tabular}{ccccc}
Tube & Length [mm] & Weight [g] & Inner diameter [mm] & Wall thickness [mm]\\ 
Short & 130 & 19 & 14 & 1\\
Long & 300 & 66 & 12 & 2 \\
\end{tabular}
\caption{Specifications of the containers.}
\label{table:test tube}
\end{center}
\end{table}

\begin{table}
\begin{center}
\begin{tabular}{ccc}
Media & Speed of sound $c$ [m/s] & Density $\rho_0$ [kg/m$^3$]\\
Silicone oil (10 cSt) & 966.5 & 935 \\
Silicone oil (1 cSt) & 901.3 & 818\\
Water & 1483 & 998\\
Ethanol & 1168 & 789\\
Gelatin & 1391 & 1052\\
\end{tabular}
\caption{Physical properties of media in a tube. Speed of sound in gelatin is calculated from the acceleration frequency and liquid depth.}
\label{table:Liquids}
\end{center}
\end{table}

\section{Results and discussion}
\subsection{Strouhal number $St$ \tb{as an indicator of liquid compressibility}}
We first derive the Strouhal number $St$ \tb{for the} one-dimensional \tb{system}. 
We assume an infinitesimal disturbance to the velocity, pressure, and density in the inviscid flow.
We then obtain the \tb{continuity and Euler equations} in terms of the density $\rho$, velocity $u$, time $t$, length $x$, and pressure $p$ as
\begin{equation}
\frac{\partial \rho}{\partial t} +\rho_{0}\frac{\partial u}{\partial x}=0,
\label{eq:renzoku2}
\end{equation}
\begin{equation}
\frac{\partial u}{\partial t}+\frac{1}{\rho_{0}}\frac{\partial p}{\partial x}=0.
\label{eq:NS2}
\end{equation}
Here, $\rho_0$ is the characteristic \tb{liquid} density in the far field.
The speed of sound in the liquid, $c$, is defined as $c^2=\partial p/\partial \rho$.
\tb{Coupling} equations (\ref{eq:renzoku2}) and (\ref{eq:NS2}) yields the wave equation \citep{Leighton1994} in the form
\begin{equation}
\frac{\partial ^{2} \Phi}{\partial t^{2}}-c^2\frac{\partial ^{2} \Phi}{\partial x^{2}}=0,
\label{eq:wave1}
\end{equation}
where $\Phi$ is the velocity potential \citep{Fujikawa2005}.
\tb{Introducing dimensionless quantities,} the wave equation can be \tb{written} as
\begin{equation}
St^2\frac{\partial^2 \Phi^{\ast}}{\partial t^{\ast2}}-\frac{\partial^2 \Phi^{\ast}}{\partial x^{\ast2}}=0,
\label{eq:wave_equation}
\end{equation}
where $St=x/(ut)$ is the Strouhal number. \tb{T}he superscript $\ast$ denotes dimensionless quantit\tb{ies}.
In this work, the length of the liquid column $L$, speed of sound $c$, and duration of acceleration $\Delta t$ are substituted for the characteristic length $x$, speed $u$, and time $t$, respectively.
The Strouhal number $St$ for our experiments \tb{is} thus defined as 
\begin{equation}
St=\frac{L}{c\Delta t},
\label{eq:St_2}
\end{equation}
which is the same form as equation (1.2). 
In this form, $St$ can be \tb{considered} as the ratio between the characteristic length of the geometry $L$ and the length scale required to develop the acoustic nature, i.e. the thickness of the wavefront of the pressure wave. 

Equation (\ref{eq:wave_equation}) also provides a theoretical insight into the fact that $St$ reflects the fluid compressibility, as reported by \cite{Reijers2017}.
For $St\ll1$, equation (\ref{eq:wave_equation}) can be rewritten as
\begin{equation}
\frac{\partial^2 \Phi^{\ast}}{\partial x^{\ast2}}=0,
\label{eq:decompression}
\end{equation}
which describes incompressible and irrotational flow, indicating that the pressure fluctuations due to wave propagation are negligible, as experimentally evidenced in figure 1. 
In contrast, for $St\not\ll 1$, the liquid compressibility and pressure fluctuations become \tb{visible} (figure 1).
The pressure fluctuations \tb{are} derived from the Euler equation (\ref{eq:NS2}). \tb{Integrating this equation} with respect to $x$ and assuming $p=p_{0}$ and ${u}=0$ in the far field gives
\begin{equation}
p-p_{0}=\rho cU,
\label{pressure}
\end{equation}
where $\rho cU$ is known as the water hammer pressure; this expression is known as Joukovski's equation \citep{Batchelor1967,Thompson1972}, a classical formulation for predicting the maximal pressure fluctuations in one-dimensional flows. 
As explained earlier, in the limit $\Delta t\approx0$, an instantaneous pressure rise is assumed and the pressure in the liquid jumps from zero to $\rho cU$.
According to the above discussion, in our experiments where the characteristic speed is $U=u_0$, the development of pressure $\overline{P}$ is expected to be \tb{dominated} by the Strouhal number $St$, which successively connects $\overline{P}=0$ at $St\ll1$ and $\overline{P}=\rho cu_0$ at $St\gg1$.

\subsection{Experimental results and remarks}
\subsubsection{Pressure fluctuations in liquids}
\begin{figure}
\begin{center}
\includegraphics[scale=0.32]{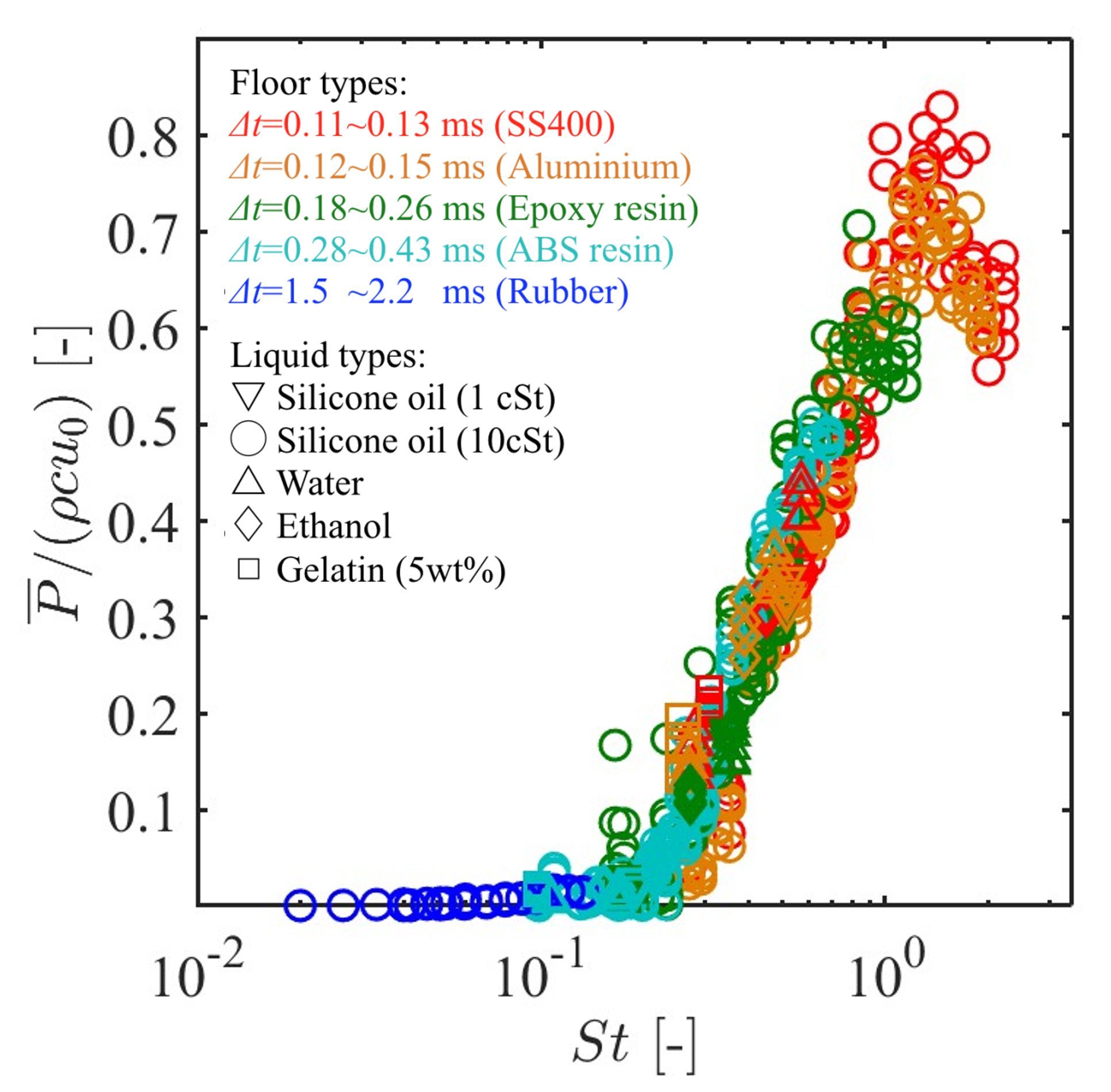}
\caption{Magnitude of pressure fluctuation $\overline{P}/(\rho c u_{0})$ versus Strouhal number $St$. Data points show experimental results and each plot colour denotes a different floor type. The shapes of the data points correspond to the different liquid types (including a hydrogel), as indicated in the legend.} 
\label{fig:Kurihara}
\end{center}
\end{figure}

Figure \ref{fig:Kurihara} compares the amplitude of the spatially averaged liquid pressure change (hereafter, the pressure fluctuation) $\overline{P}$ to the Strouhal number {\it St} for different values of $\Delta t$ and $L$.
The use of different floor materials \tb{results in} $\Delta t$ varying from 0.1--2.2~ms, as indicated in the legend. 
The use of water ($\bigtriangleup$), 1 cSt silicone oil ($\bigtriangledown$), and ethanol ($\diamond$) allows us to vary the speed of sound $c$ and the liquid density $\rho$, which are related to the water hammer pressure.
In particular, the speed of sound $c$ in water is approximately 1.6 times \tb{faster than} that in 1 cSt silicone oil. The density $\rho$ of water is approximately 1.3 times \tb{higher than} that of ethanol.

The experimental data fall onto a single curve that is insensitive to the experimental parameters \tb{tested}.
All of the data for $St>0.2$ overlap significantly and approach $\overline{P}/(\rho c u_{0})\approx1.0$ when $St\geq O(1)$, \tb{whereas the data using resin floors for $St\leq$0.2 are scattered}.
\tb{The collapse of the data in figure \ref{fig:Kurihara} indicates that the Strouhal number $St$ describes the gradual pressure development in a one-dimensional tube. Of the conditions tested, $St\sim0.2$ is the threshold at which the pressure fluctuations become visible and the transition begins ($\overline{P}/(\rho c u_{0})\approx0.1$).}

Figure \ref{fig:Kurihara} also contains data for 10 cSt silicone oil ($\bigcirc$).
This has a greater viscosity than the other liquids, but the same trend can be observed because the viscous boundary layer $\delta\sim\sqrt{\nu\Delta t}\sim O(10^{-2})$~mm is expected to be sufficiently thinner than the container radius $\sim O(1)$ mm, as argued in the case of jetting experiments \citep{Onuki2018,Gordillo2020}. 
The surface tension does not significantly affect the pressure fluctuations in the present experiments, despite the surface tension of water being approximately 4.3 times \tb{higher than that of} 1~cSt silicone oil.
\tb{From the above, the Strouhal number $St$ provides a powerful means of describing the pressure fluctuations inside widely used low-viscosity liquids.}

Previous research on similar systems suggests that the accelerating fluid can be modelled as either a compressible fluid \citep{Fogg2009,Kiyama2016,Daou2017,Yukisada2018,Kamamoto2021b} or an incompressible fluid \citep{Jacobs2003,Taylor2012,Daily2014,Fatjo2016,Pan2017,Gordillo2020,Kamamoto2021,Krishnan2022}, even if the Strouhal number is in the intermediate region $St\approx O(1)$.
Our experimental results (figure \ref{fig:Kurihara}) obviously show that the pressure wave appears at moderate values of $St$ and its magnitude scales with $St$: the transition from the ``incompressible fluid" region starts at $St\sim O(10^{-1})$ \tb{($St\approx0.2$ in this experiment)} and the ``compressible fluid" region is established at $St\sim O(10^{1})$.
This is reasonably consistent with the case of laser-ablation onto a droplet \citep{Reijers2017}, in which fluid compressibility appears in the flow field (and accordingly the pressure field) at $St\gg1$.

\subsubsection{Cavitation inception following pressure fluctuations}
\tb{It is thus expected that the Strouhal number $St$ could provide a measure of the tendency for cavitation in a one-dimensional fluid system. The cavitation number for the accelerated liquid \citep{Pan2017} is written as}

\begin{equation}
Ca=\frac{P_{atm}-P_v}{\rho aL},
\end{equation}
\tb{where $a=\overline{U}/\Delta t$. Employing the first-order approximation $\rho L(\overline{U}/\Delta t)\sim St\rho cu_0$, we obtain}
\begin{equation}
Ca\sim St^{-1}\frac{P_{atm}-P_v}{\rho c\overline{U}}.
\end{equation}
\tb{This expression makes sense for situations in which the influence of $\Delta t$ is dominant over that of other parameters. If $\Delta t$ is dominant, $Ca$ and $St^{-1}$ play a similar role and are thus interchangeable. Increasing $St$ decreases $Ca$, meaning a greater pressure reduction, as expected, and a higher possibility of cavitation occurring \citep{Daily2014,Pan2017,Eshraghi2022}.}
This perhaps explains the findings in previous studies \citep{Fatjo2016,Pan2017,Xu2021,Zhichao2022}, in which the overall cavitation tendency was reasonably classified by only considering $Ca$ as the significance of $\Delta t$ remained largely unchanged.

\begin{figure}
\begin{center}
\includegraphics[width=0.8\columnwidth]{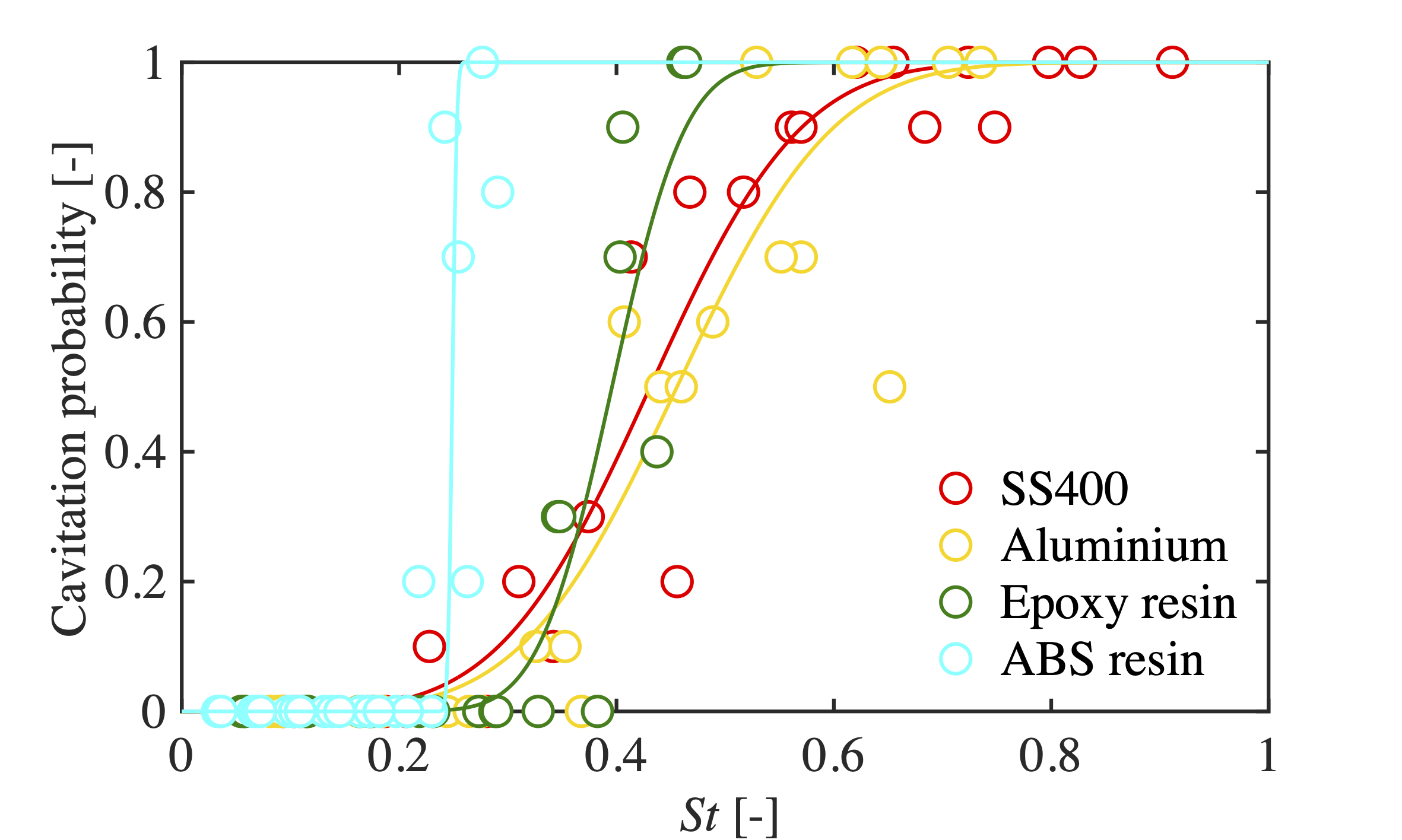}
\caption{Cavitation probability of silicone oil (10 cSt) for various floor materials, as indicated by colours. The probabilities were experimentally determined through 10 separate experimental runs. The fitting curves were obtained based on equation (3.10), where the fitting parameters are $\alpha=0.431$ and $\beta=0.109$ for SS400, $\alpha=0.456$ and $\beta=0.113$ for aluminium, $\alpha=0.396$ and $\beta=0.053$ for epoxy resin, and $\alpha=0.249$ and $\beta=0.0032$ for ABS resin.}
\label{fig:Cavitation}
\end{center}
\end{figure}

\tb{Figure \ref{fig:Cavitation} shows the probability curves of cavitation in silicone oil (10 cSt) for different floor types. Each marker shows the cavitation probability determined through 10 separate experimental runs. The gradual increase in probability is fitted by the sigmoid function \citep{Maxwell2013,Hayasaka2017,Bustamante2017,Oguri2018,Bustamante2019} as}

\begin{equation}
Prob.=\frac{1}{2}\bigg[1+\mathrm{erf}\bigg(\frac{St - \alpha}{\beta\sqrt{2}}\bigg)\bigg],
\end{equation}

\noindent \tb{where $\mathrm{erf()}$ is the error function implemented in Matlab, and $\alpha$ and $\beta$ are the fitting parameters. Overall, cavitation starts at $St$ values as low as 0.2, which agrees well with the condition for noticeable pressure fluctuations. Closer observation reveals the influence of the floor material. For relatively stiff floors (i.e. SS400 and aluminium), a 50\% probability of cavitation is reached at $St=0.43$ and $St=0.46$, respectively. For softer floors, the probability curve is steeper (i.e. smaller $\beta$ values), and the minimum $St$ value required does not vary significantly. The physical reasoning behind this remains unclear, and further investigation is needed. The $\Delta t$ value varies for softer floors (figure \ref{fig:Kurihara}), so its influence on the results needs to be clarified.}

\subsubsection{Pressure fluctuations and cavitation in a hydrogel}
\tb{The Strouhal number $St$ is still a powerful tool for describing the pressure fluctuations in hydrogels. In figure \ref{fig:Kurihara}, the gelatin data (squares) for various floor types are in good agreement with the other data for standard fluids.}
The 5 wt\% gelatin gel is a water-based viscoelastic material, which behaves as a soft solid in the rest state.
At the moment of impact, it is expected that the gelatin gel could flow due to the fast deformation ($\Delta t\sim O(0.1)$~ms). 
\tb{The above discussion on the Strouhal number $St$ and the water hammer pressure $\rho cu_0$} can be applied to the accelerated hydrogel, although \tb{further investigations are needed for} a quantitative \tb{understanding}. 
Our findings provide an \tb{experimental} understanding of the inception of cavitation inside the gel \citep{Kang2017,Kang2017b,Kang2018}; see the supplementary materials. \tb{This phenomenon} has been used to address brain injury issues \citep{Yu2020,Lang2021}.

\subsection{\tb{Analysis of the pressure fluctuations}}
{We now discuss the basic theoretical \tb{framework} for predicting the pressure fluctuations.
First, we consider a \tb{gradual} pressure \tb{increment at the pressure wavefront} to confirm the smooth connection between the ``incompressible fluid" region ($St\ll1$) and the ``compressible fluid" region ($St>1$) as a function of the Strouhal number $St$ (\S 3.3.1).
We then compare the theoretical models and the experimental data (\S 3.3.2).} 
\tb{We discuss the influence of both the container mass and the pressure wavefront profile.}

\subsubsection{Model considering thickness of pressure wave}
{This paper focuses on the case of $St=L/(c\Delta t)\approx1$, where the length of the liquid column $L$ is similar to the length $c\Delta t$.
We call $c\Delta t$ the thickness of the pressure wavefront because it \tb{corresponds to the} length required to establish the water hammer pressure. 
In such cases, the pressure fluctuations are not determined by the classical formulation assuming an instantaneous pressure rise [equation (\ref{pressure})], but are caused by both the temporal pressure change during $\Delta t$ and the reflection of the pressure wave at the boundaries.}

\begin{figure}
\begin{center}
\includegraphics[scale=0.14]{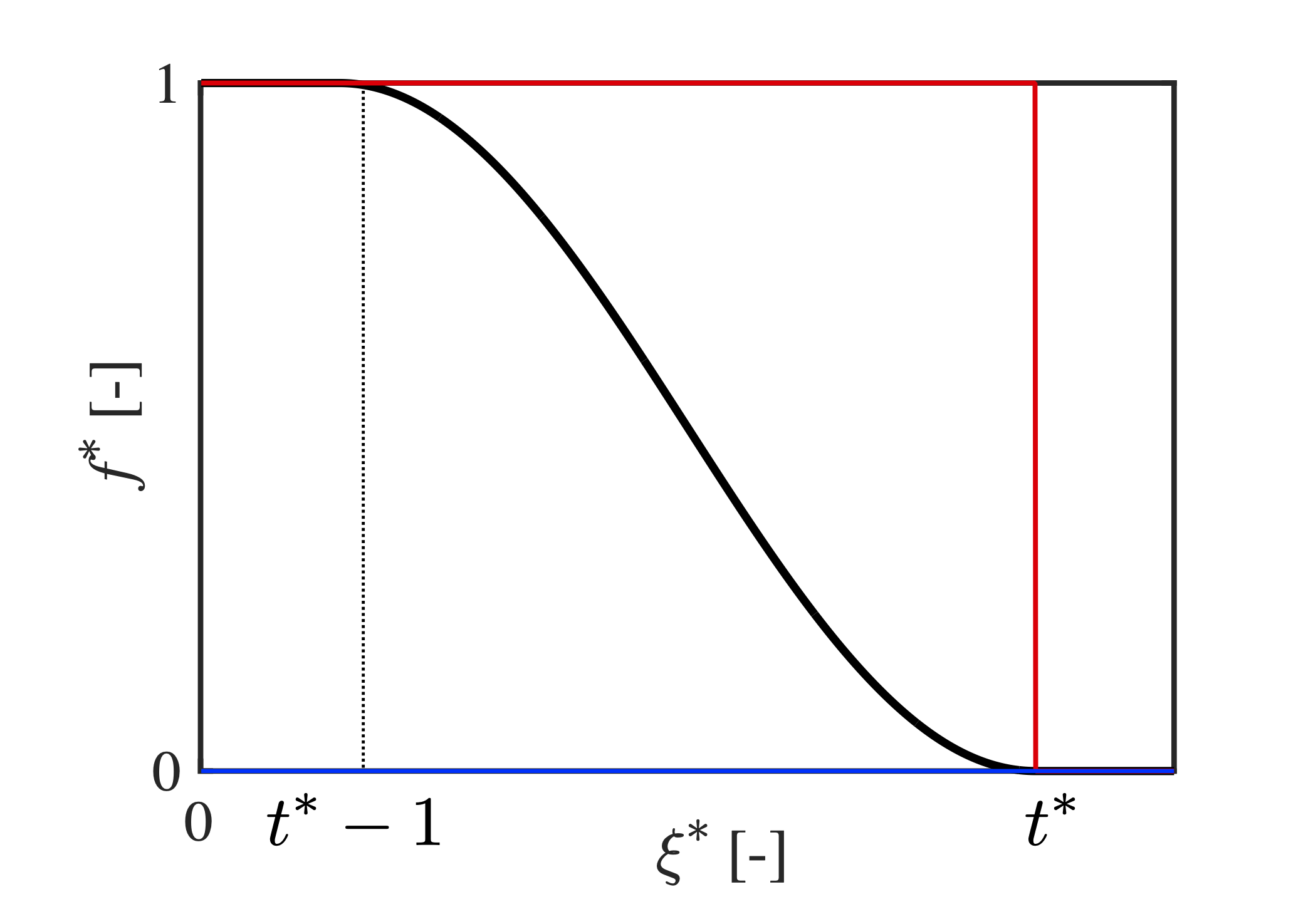}
\caption{Outline view of the function of pressure change $f^{\ast}$, described by the dimensionless time $t^{\ast}$ and dimensionless position $\xi^{\ast}$. Red, blue, and black lines show models with the assumption that the thickness of the wavefront is 0 (i.e. water hammer theory), that the liquid is incompressible, and that the wavefront has a finite thickness, as modelled herein, respectively.}
\label{fig:model1}
\end{center}
\end{figure}

The liquid column in the container has a gas--liquid interface at one end and a solid--liquid interface at the other end.
First, we define the function $f^{\ast}(\xi^{\ast},t^{\ast})$ describing the liquid pressure change, where $\xi^{\ast}$ and $t^{\ast}$ are the distance from the solid--liquid interface and the time from the initial impact, respectively.
The pressure change propagates in the direction of $\xi^{\ast}$.
The wavefront is located at $\xi^{\ast}=t^{\ast}$ at $t^{\ast}$.
We express $f^{\ast}(\xi^{*},t^{*})$ \tb{as}
\begin{equation}
f^{\ast}(\xi^{\ast},t^{\ast}) = \left \{
\begin{array}{lll}
1 & (t^{\ast}-1 \geq \xi^{\ast} )\\
I^{\ast}(\xi^{\ast}-t^{\ast}+1) & (t^{\ast}-1 \leq \xi^{\ast} \leq t^{\ast} )\\
0 & (t^{\ast} \leq \xi^{\ast}),
\end{array}\right.
\label{eq:wave_surface}
\end{equation}
where $I^{\ast}(\eta)$ is the pressure change in the pressure wavefront, \tb{which is} modelled as
\begin{equation}
I^{\ast}(\eta) =\frac{1}{2}\sin(\eta\pi+\frac{\pi}{2})+\frac{1}{2};
\end{equation}
\tb{see also figure \ref{fig:model1}.}
The reflection of the pressure wave at the boundaries is modelled by assuming the principle of superposition \citep{Thompson1972}.
{We assume that the mismatching of the acoustic impedance at the gas--liquid and solid--liquid interfaces is significant, and hence we neglect the energy loss upon reflection.}
The sign of the pressure reverses when the pressure wave reflects at the gas--liquid interface, but does not change when the pressure wave collides the solid--liquid interface.
The local pressure in the liquid column $p^{\ast}$ is calculated as follows:

\begin{equation}
p^{\ast}(\xi^{\ast},t^{\ast},St)=\sum_{i=1}^{ceil(\frac{t^{\ast}}{St})}g^{\ast}(\xi^{\ast},t^{\ast},St,i),
\label{eq:cases}
\end{equation}
\[
g^{\ast}(\xi^{\ast},t^{\ast},St,k)=\left \{
\begin{array}{llll}
f^{\ast}(kSt-\xi^{\ast},t^{\ast})&(k=4n)\\
-f^{\ast}((k-1)St+\xi^{\ast},t^{\ast})&(k=4n-1)\\
-f^{\ast}(kSt-\xi^{\ast},t^{\ast})&(k=4n-2)\\
f^{\ast}((k-1)St+\xi^{\ast},t^{\ast})&(k=4n-3),\\
\end{array}\right.
\]
where $n$ is a natural number.

This model allows us to estimate the temporal changes in pressure.
We then calculate the amplitude of the spatially averaged pressure in the liquid, $\overline{P}$.
Hereafter, the amplitude of the pressure change $\overline{P}/(\rho cu_0)$ calculated from the above model is compared with that estimated from the experiments, $mV/(\rho ALu_0)$.

\subsubsection{Comparison and discussion}
\begin{figure}
\begin{center}
\includegraphics[scale=0.23]{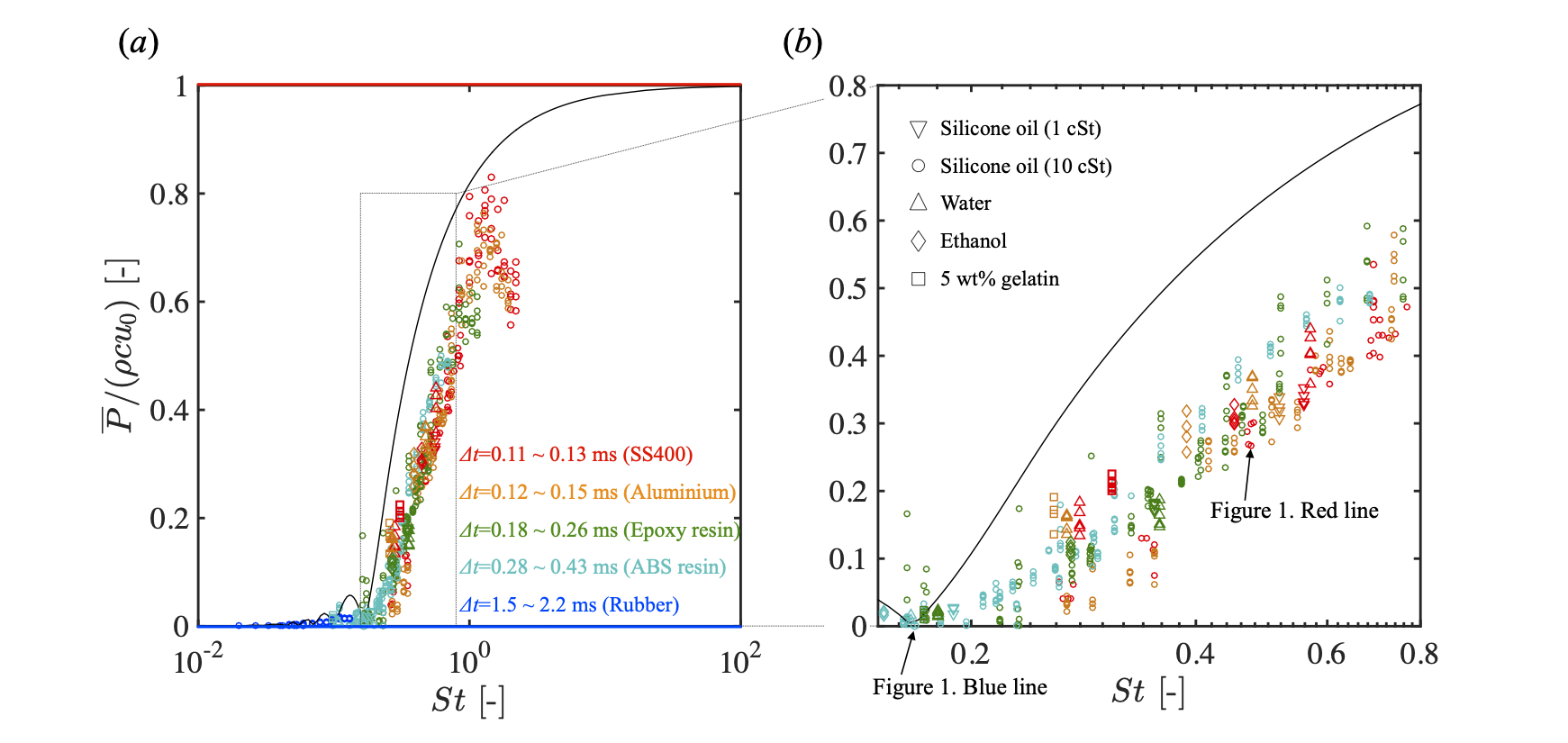}
\caption{Dimensionless amplitude of the spatial pressure in a liquid as a function of the Strouhal number, $St$. Data points show experimental results and each plot colour corresponds to a different floor type. The shapes of the data points correspond to different liquids. The red and blue lines show the pressure calculated under the assumptions compressible and incompressible liquids, respectively. The black thin line represents the calculation result using the model (\S 3.3.1).} 
\label{fig:Main_figure}
\end{center}
\end{figure}

{We now compare the experimental results (dataset visualised in figure \ref{fig:Kurihara}) with the model in figure \ref{fig:Main_figure}({\it a}).
The black solid line expresses our model with consideration of the pressure wavefront thickness (see \S 3.3.1).
Figure \ref{fig:Main_figure}({\it b}) shows a magnified view.
For $St = O(10^{-2})$, the fluctuations are not obvious in either the model or the experimental results (${\overline{P}}/({\rho c u_{0}})<10^{-2}$), suggesting that the liquid behaves as an incompressible fluid.
For $St = O(10^{-1})$, the fluctuations in both the experimental results and the model increase notably, which is associated with an increase in the Strouhal number $St$. 
Therefore, the effect of liquid compressibility appears at $St = O(10^{-1})$.}
The proposed model predicts the onset of pressure fluctuations (see figure \ref{fig:Main_figure}{\it b}, $St\sim0.2$) and captures the overall trend of these fluctuations with respect to $St$, but overestimates the magnitude of the fluctuations for higher $St$ values ($St<O(10^{0})$).
\tb{Though the model implies that the water hammer pressure (${\overline{P}}/({\rho c u_{0}})=1.0$) is reached at higher $St$ values,} the experimental results saturate at around ${\overline{P}}/({\rho c u_{0}})\sim0.7$ for $St\sim1$.
Although no data could be obtained for $St>O(10^{1})$ due to experimental limitations, we expect that the experimental scenario would differ for higher $St$ values as well \tb{because of the influence of the container motion and the profile for the pressure wavefront, as discussed below}.

First, we investigate the effects of the container motion.
The momentum conservation law for a system consisting of a container and a liquid is 
\begin{equation}
mv+A\rho\int_{0}^{L} udx=H(const.),
\label{momentum_all}
\end{equation}
where $m$ is the mass of the container, $v$ is the velocity of the container, and $A$ is the cross-sectional area of the liquid column.
Note that we neglect the contribution of gravitational acceleration because it is expected to be very small when compared with the acceleration imposed by the impact ($g\ll u_0/\Delta t$). 
Equation (\ref{momentum_all}) can be reformulated as follows in dimensionless quantities:
\begin{equation}
mv^{\ast}u_{0}+A\rho u_{0} c \Delta t\int_{0}^{St} u^{\ast}dx^{\ast}=H(const.).
\label{v_dimensionless}
\end{equation}
The momentum conserved in the system, $H$, is calculated using $v^{\ast}$ and $u^{\ast}$ immediately after the impact.
The velocity of the container $v^{\ast}$ is then calculated from the constant momentum $H$ and the instantaneous velocity in the liquid $u^{\ast}$.
We incorporate the function $f^{\ast}$ in {\S} 3.3.1 into the velocity of the container, $v^{\ast}$, as
\begin{equation}
f^{\ast}(\xi^{\ast},t^{\ast})=\left \{
\begin{array}{lll}
v^{\ast} & (t^{\ast}-1 \geq \xi^{\ast} )\\
I^{\ast}(\xi^{\ast}-t^{\ast}+1) & (t^{\ast}-1 \leq \xi^{\ast} \leq t^{\ast} )\\
0 & (t^{\ast} \leq \xi^{\ast}).
\end{array}\right.
\end{equation}
The pressure in the liquid is calculated from equation (\ref{eq:cases}) using the function $f^{\ast}$.
The velocity of the liquid, $u^{\ast}$, is calculated from the following equation:
\begin{equation}
u^{\ast}(\xi^{\ast},t^{\ast},St)=\sum_{i=1}^{ceil(\frac{t^{\ast}}{St})}h^{\ast}(\xi^{\ast},t^{\ast},St,i),
\label{eq:velocity_calculation}
\end{equation}
\[
h^{\ast}(\xi^{\ast},t^{\ast},St,k)=\left \{
\begin{array}{llll}
-f^{\ast}(kSt-\xi^{\ast},t^{\ast})&(k=4n)\\
-f^{\ast}((k-1)St+\xi^{\ast},t^{\ast})&(k=4n-1)\\
f^{\ast}(kSt-\xi^{\ast},t^{\ast})&(k=4n-2)\\
f^{\ast}((k-1)St+\xi^{\ast},t^{\ast})&(k=4n-3).\\
\end{array}\right.
\]

The velocity of the container, $v^{\ast}$, is dependent on the momentum of the liquid, {which} implies some dependence on $L$, $m$, $A$, and $\rho$ [equation (\ref{v_dimensionless})], \tb{and can be interchanged with the pressure, as we did for the experimental data}.
Considering the velocity change of the container, the amplitude of the pressure fluctuations in the liquid {is} not only defined by the Strouhal number $St$ and the water hammer pressure.

\begin{figure}
\begin{center}
\includegraphics[scale=0.105]{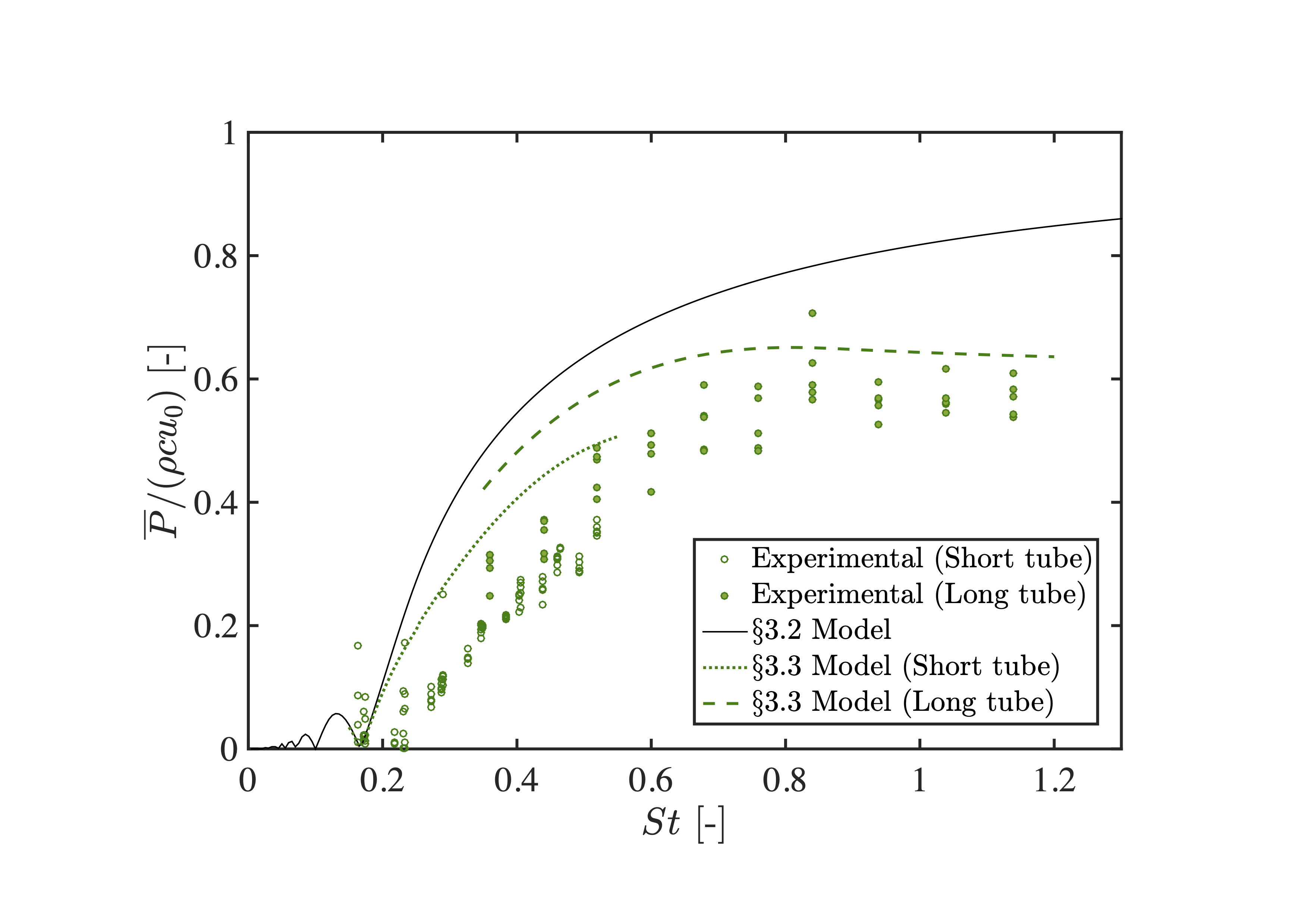}
\caption{Effect of container motion on the pressure fluctuations in a liquid. Open (filled) circles show experimental results using short (long) tube. The dotted (dashed) line shows the model results considering the motion of the short (long) container.}
\label{fig:Sub_figure}
\end{center}
\end{figure}

The \tb{revised model} is compared with the experimental results in the case of an epoxy resin floor (figure \ref{fig:Sub_figure}).
\tb{We used two different tubes to vary the mass;} the filled and open circles represent the data for the long and short tubes, respectively.
The dashed and dotted lines correspond to the model for the long and short tubes, respectively.
The revised model taking the change in container velocity into account improves accuracy in terms of the overestimation observed in the original models. The revised model for different tubes also predicts the influence of the tube types, whereas such variations might be not necessarily visible in the experimental data.

\begin{figure}
\begin{center}
\includegraphics[scale=0.24]{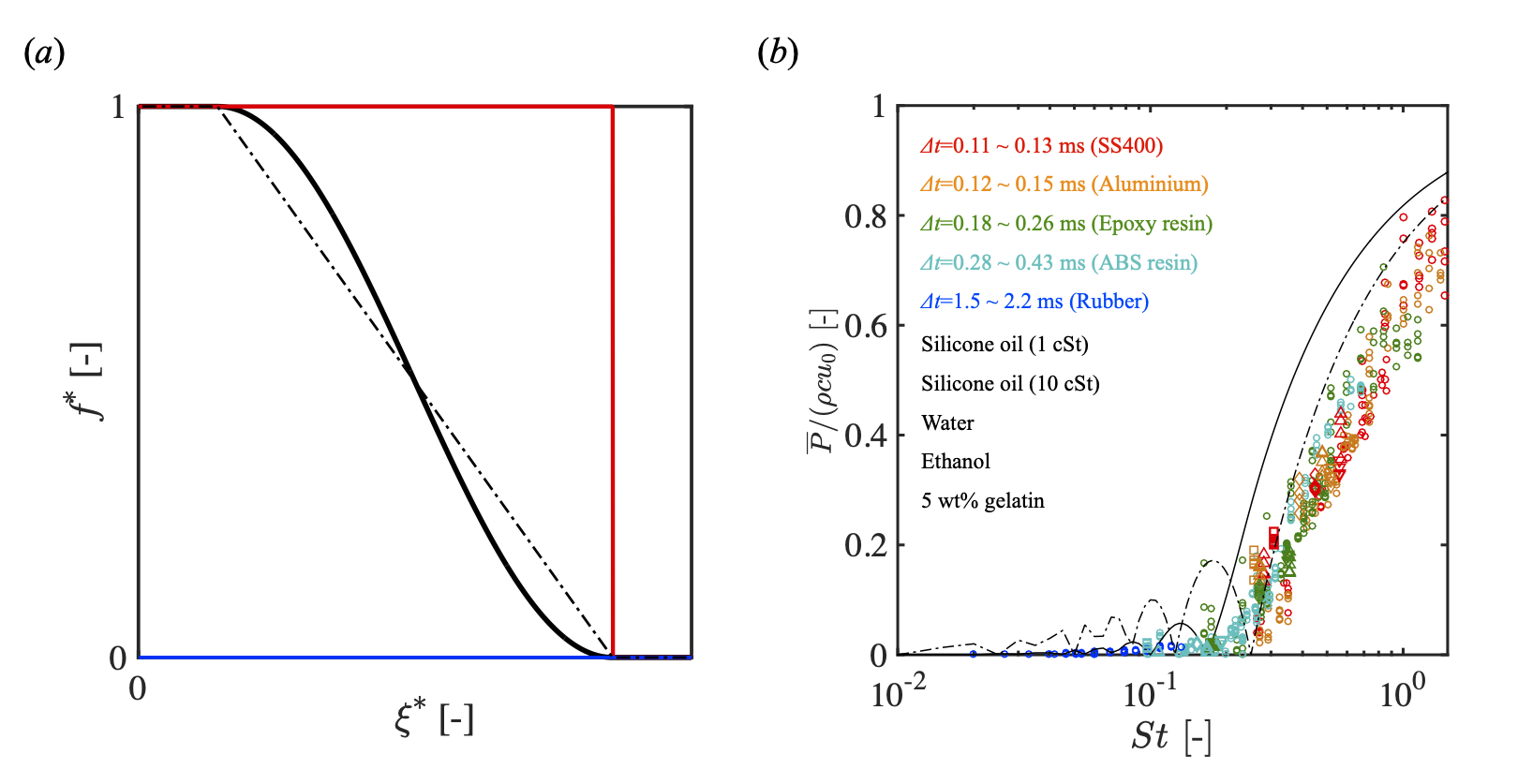}
\caption{(a) Profile of pressure wavefront. (b) Dimensionless fluctuations in water hammer pressure with respect to Strouhal number. The solid and dot-dashed lines represent different wavefront profiles.}
\label{fig:main_liner}
\end{center}
\end{figure}

We \tb{also} tested the effect of the pressure wavefront profile.
For comparison purposes, we \tb{used a} simple profile, illustrated by the dot-dashed line in figure \ref{fig:main_liner}({\it a}).
This is expressed by $I^{\ast}(\eta)$ in equation (\ref{eq:wave_surface}), which is defined as 
\begin{equation}
I^{\ast}(\eta)=1-\eta. 
\end{equation}
This model overpredicts the pressure fluctuations for smaller $St$ values ($St<O(10^{-1})$), \tb{but gives better agreement for larger $St$ values (figure \ref{fig:main_liner}{\it b}).}
This suggests that the profile of the pressure wavefront could affect the accuracy of the model \tb{predictions}, which provides a possible explanation for the difference between the experimental results and the original model (figure \ref{fig:Main_figure}).
In our experiments, the profile of the pressure wavefront is assumed to be dependent on the floor material.
Hence, the effect of the pressure profile causes variations in the pressure fluctuations measured experimentally at the same $St$ \tb{as well as the cavitation trend for each floor type}.

In addition to the above considerations, \tb{there are two possible reasons for differences between the model and experiments: the vibration of the container and the curvature of the bottom of the container.}
Although the model does not consider the effect of vibrations in the container, this phenomenon was indeed observed in the experiments (see figure \ref{fig:Acceleration}{\it a}). The momentum in the experimental system is converted to vibrations. The vibration of the container after the rebound might also be related to the amount of liquid inside the system \citep{Taylor2012,Kiyama2016,Klebbert2023}.
Additionally, the model does not consider the curvature of the bottom of the container, which leads \tb{to} the \tb{formation} of a plane wave in this region.
\tb{However}, we used a container \tb{with a} rounded bottom in the experiments.
A radial phase lag is likely to occur and the fluctuations will differ from those of the model.

\section{Conclusions}
This paper has shown that the Strouhal number $St$ provides a measure of the fluid compressibility in a one-dimensional system, unlike other dimensionless indicators such as $Ma$ and $Fr$.
Experiments with various liquid column depths inside a tube were performed at different liquid velocities, acceleration durations, and media types (four standard liquids and one hydrogel), producing a wide range of Strouhal numbers, $0.02 \leq St \leq 2.2$.
\tb{The pressure fluctuations, which were calculated based on the measured acceleration, indicate a unified trend as a function of $St$} for a wide range of parameters (figure \ref{fig:Kurihara}), suggesting that $St$ gives a good measure of the liquid compressibility in cases where the pressure impulse approach breaks down.
Our experimental findings regarding the importance of $St$ are consistent with those from numerical work on bursting droplets \citep{Reijers2017}, suggesting that our discussion related to $St$ might even be applicable in different geometries.
\tb{We also found that $St$ is useful for describing the cavitation tendency (figure \ref{fig:Cavitation}).}
In addition to empirical relations, we derived a conceptual model in which a modified version of water hammer theory accounts for the finite thickness of the pressure wavefront.
\tb{This} simple model describes the overall trend of the experimental data: the magnitude of the pressure fluctuations increases as the Strouhal number $St$ increases (figure \ref{fig:Main_figure}). 
Implementing corrections in the model allowed us to show that both the motion of the surrounding container and the profile of the pressure wavefront influence the pressure development (see figures \ref{fig:Sub_figure} and \ref{fig:main_liner}).

\section*{Acknowledgements}
We thank Prof. J. Rodr\'{i}guez-Rodr\'{i}guez for fruitful discussions.
We also thank Dr. Andres Franco-Gomez, Prof. Masakazu Muto, Dr. Hajime Onuki, Dr. Prasad Sonar from Tokyo University of Agriculture and Technology for proofreading the manuscript and making a number of helpful suggestions.
This work was partly supported by JSPS KAKENHI through grant numbers JP16J08521, JP17H01246, JP20H00223, JP20H00222, JP20K20972, and JP23K19089, the Japan Science and Technology Agency PRESTO (Grant No. JPMJPR21O5), and Japan Agency for Medical Research and Development (Grant No. JP22he0422016) and by funding from the Institute of Global Innovation Research at the Tokyo University of Agriculture and Technology.

\bibliography{Kurihara_LP}
\bibliographystyle{jfm}

\section*{Supplementary materials}
\subsection*{Cavitation in a hydrogel system}
\begin{figure}
\begin{center}
\includegraphics[width=0.8\columnwidth]{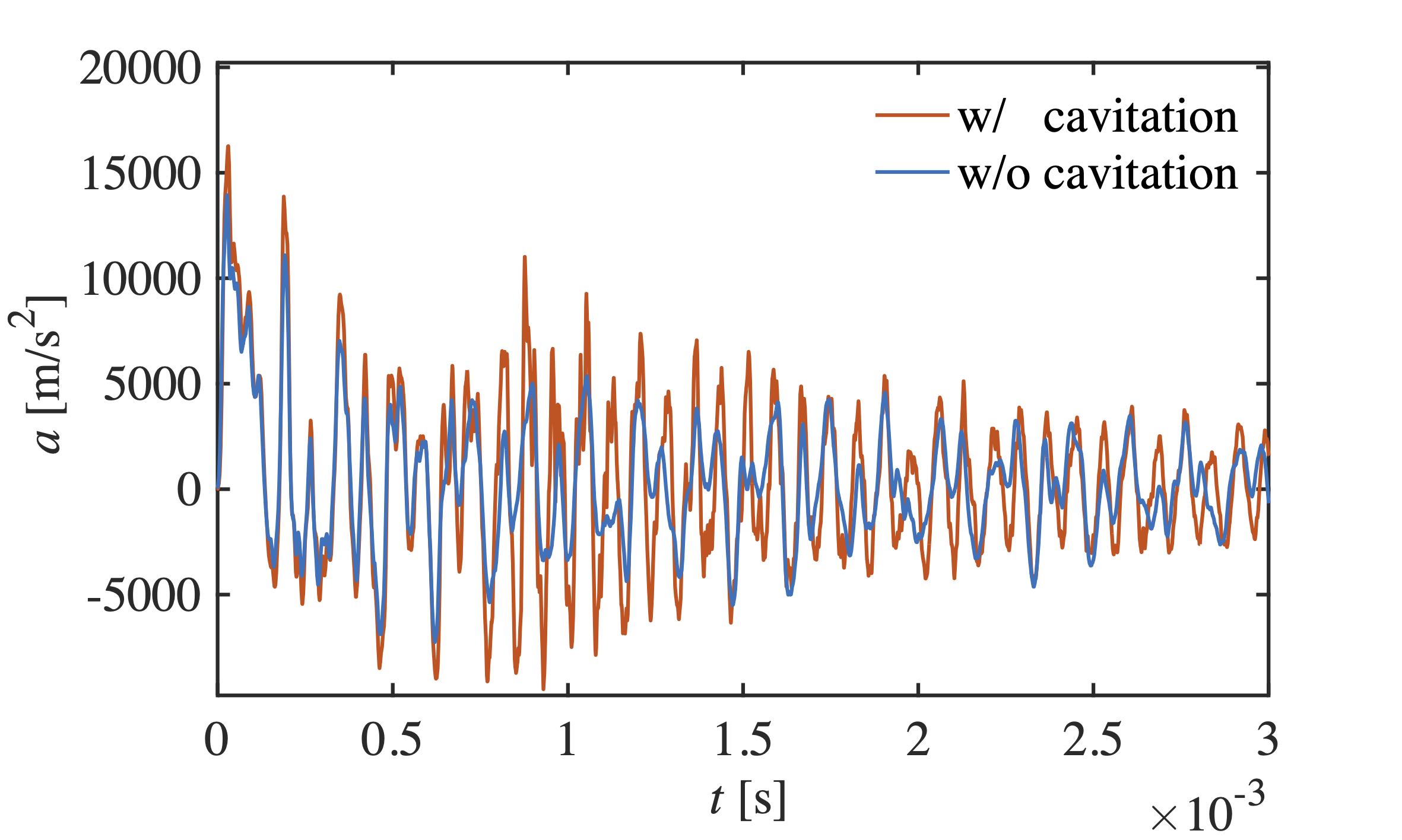}
\caption{Acceleration $a$~(m/s$^2$) measured as a function of time $t$~(s). The orange line represents the data with cavitation, while the blue line shows data without cavitation onset. The liquid column height and drop height are $L=60$~mm and $H=15$~mm. The Strouhal number $St$ is expected to be greater than 0.2. The signals have not been low-pass filtered to avoid the elimination of high-frequency responses caused by bubble activity.}
\label{fig:Gelatine}
\end{center}
\end{figure}

The onset of cavitation and the associated dynamics are of great interest in the community, as they contribute to understanding the brain injuries that may occur following a sudden impact \citep{Barney2020,Marsh2021}. Though a similar system observed the onset of cavitation in a gelatin hydrogel using a high-speed camera \citep{Pan2017}, the present study did not perform any direct visualizations. However, comparing acceleration data as a function of time, some cases exhibit enhanced signals at $O(1)$~ms after the impact (see figure \ref{fig:Gelatine}). These might be related to the perturbation inside the gelatin column, namely the onset of cavitation. This result supports our claim that the present research has the potential to clarify our understanding of acceleration-induced brain injuries, although the sensitivity of measurements is limited.

\end{document}